\documentclass[a4paper]{article}

\usepackage{INTERSPEECH2018}
\usepackage{microtype}
\usepackage{booktabs}
\usepackage{amsmath}

\DeclareMathOperator{\Tr}{Tr}

\title{Improved MVDR Beamforming Using LSTM Speech Models to Clean Spatial Clustering Masks}
\name{Zhaoheng Ni$^1$, Felix Grezes$^1$, Viet Anh Trinh$^1$, Michael I.~Mandel$^{12}$}
\address{
  $^1$Computer Science Program, The Graduate Center, City University of New York\\
  $^2$Computer and Information Science, Brooklyn College, City University of New York}
\email{\{zni,fgrezes,vtrinh\}@gradcenter.cuny.edu, mim@sci.brooklyn.cuny.edu}

\begin{document}

\maketitle
\begin{abstract}

Spatial clustering techniques can achieve significant multi-channel noise reduction across relatively arbitrary microphone configurations, but have difficulty incorporating a detailed speech/noise model. In contrast, LSTM neural networks have successfully been trained to recognize speech from noise on single-channel inputs, but have difficulty taking full advantage of the information in multi-channel recordings.
This paper integrates these two approaches, training LSTM speech models to clean the masks generated by the Model-based EM Source Separation and Localization (MESSL) spatial clustering method.  By doing so, it attains both the spatial separation performance and generality of multi-channel spatial clustering and the signal modeling performance of multiple parallel single-channel LSTM speech enhancers. 
Our experiments show that when our system is applied to the CHiME-3 dataset of noisy tablet recordings, it increases speech quality as measured by the Perceptual Evaluation of Speech Quality (PESQ) algorithm and reduces the word error rate of the baseline CHiME-3 speech recognizer, as compared to the default BeamformIt beamformer.

\end{abstract}


\section{Introduction}
\label{sec:intro}
With speech recognition techniques approaching human performance on noise-free audio with a close-talking microphone \cite{achieving-human-parity-conversational-speech-recognition-2}, recent research has focused on the more difficult task of speech recognition in far-field, noisy environments. This task requires robust speech enhancement capabilities.

One approach to speech enhancement is spatial clustering, which groups together spectrogram points coming from the same spatial location \cite{MandelEtAl2017}.  This information can be used to drive beamforming, which linearly combines multiple microphone channels into an estimate of the original signal that is optimal under some test-time criterion \cite{brandstein2013microphone}.  This optimality is typically based on properties of the signals or the spatial configuration of the recordings at test time, with no training ahead of time.

Another approach is to use speech/noise signal models trained from data. Recent work on deep recurrent neural networks using the LSTM architecture \cite{hochreiter1997long} can achieve significant single-channel noise reduction \cite{6638947, weninger2015speech}, and so there is interest in using trainable deep-learning models to perform beamforming. This is especially useful for optimizing beamformers directly for automatic speech recognition \cite{xiao16, SainathEtAl2015}, although such optimization must happen at training time on a large corpus of training data.  Such models have difficulty generalizing across microphone arrays, especially when the number of microphones and the geometry of the array can vary.

In contrast to deep learning-based beamforming, spatial clustering is an unsupervised method for performing source separation, so it easily adapts across microphone arrays \cite{5200357, sawada2011underdetermined, bagchi15}.  Such methods group spectrogram points based on similarities in spatial properties, but are typically not able to take advantage of signal models, such as models of speech or noise. 

Developed by Mandel et al. \cite{5200357}, Model-based EM Source Separation and Localization (MESSL) is a system that computes time-frequency spectrogram masks for source separation as a byproduct of estimating the spatial location of the sources. It does so using the expectation maximization (EM) algorithm, iteratively refining the estimates of the spatial parameters of the audio sources and the spectrogram regions dominated by each source.

While MESSL utilizes spatial information to separate multiple sources, it does not model the content of the signals themselves.  This is an advantage when separating unknown sources, but performance can be improved when a model of the target source is available. The goal of this paper is to improve MESSL's performance when applied to noisy speech. We do this by training an LSTM neural network to clean the spatial clustering masks produced by MESSL.

In this paper we describe a novel method of combining single-channel LSTM-based speech enhancement into the MESSL spatial clustering system. We train a separate LSTM model that uses the single-channel noisy audio to clean the masks produced by MESSL.
Our results are evaluated using the CHiME-3 dataset, consisting of 6-channel audio recordings simulating the use of a tablet in noisy environments.  Performance is measured in terms of audio quality, as measured by the PESQ algorithm \cite{rix2001perceptual} as well as the word error rate (WER) of the baseline CHiME-3 recognizer implemented in the Kaldi speech recognition toolkit \cite{Povey_ASRU2011} applied to the enhanced signals.  It is compared to the baseline of MESSL-only \cite{mandel16b} and signals enhanced by the BeamformIt \cite{anguera2006beamformit} delay and sum beamformer.
Our method improves the PESQ score by 0.60 and the WER by 3.3\% (absolute) when compared to the BeamformIt baseline.

\begin{figure*}[ht!]
    \centering
    \includegraphics[width=0.95\textwidth]{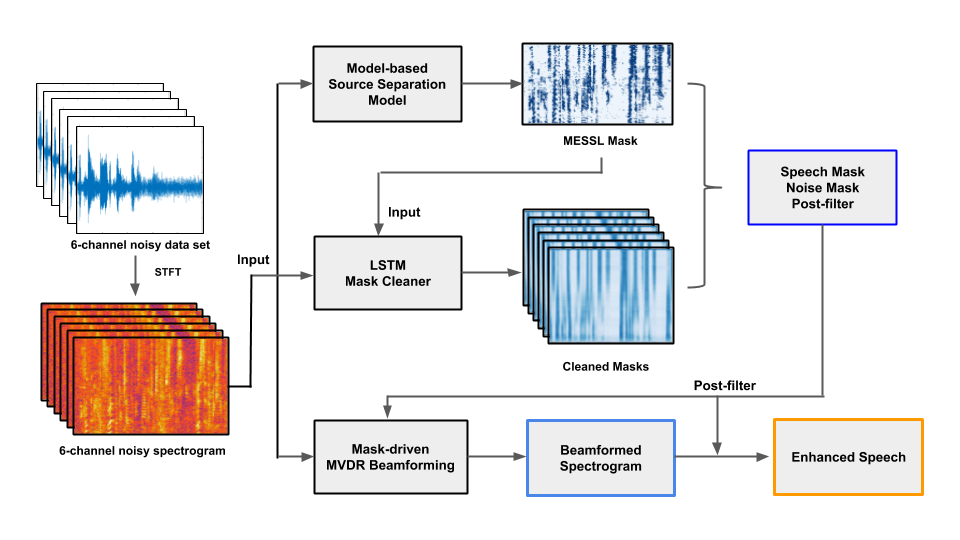}
    \caption{Multi-channel Spatial Clustering Based Time-Frequency Mask Enhancement System}
    \label{fig:framework}
\end{figure*}

\section{Related Work}
\label{sec:related}
Recently, Nugraha et al. \cite{nugraha2016multichannel} also studied multi-channel source separation using deep feedforward neural networks, using a multi-channel Gaussian model to combine the source spectrograms, to take advantage of the spatial information present in the microphone array. They explore the efficacy of different loss functions and other model hyper-parameters. One of their findings is that the standard mean-square error loss function performed close to the best. In contrast to our work they do not use spatial information that beamforming can give.

Pfeifenberger et al. \cite{pfeifenberger2017} proposed an optimal multi-channel filter which relies solely on speech presence probability. This speech-noise mask is predicted using a 2-layer feedforward neural network using features based on the leading eigenvector of the spatial covariance matrix of short time segments. Using a single eigenvector makes the input to the network independent of the number of microphones, and thus adaptable to new microphone configurations.  It is trained on the simulated noisy data portion of CHIME-3. They show that this filter improves the PESQ score of the audio.  This approach uses an early fusion of the microphone channels before they are processed by the network, as opposed to our late fusion after the network processes each channel.

Heymann et al \cite{heymann2016neural,heymann2017} also study the combination of multi-channel beamforming with single-channel neural network model. Similar to ours, the proposed model consists of a bidirectional LSTM layer, followed by feedforward layers, in their case three. Of particular note is the companion paper by Boeddeker et al. \cite{boeddeker2017}, which derives the derivative of an eigenvalue problem involving complex-valued eigenvectors, allowing their system to propagate errors in the final SNR through the beamforming and back to the single-channel DNNs.  While we do not optimize our system in this end-to-end manner, the combination of MESSL with the per-channel DNNs may provide advantages in modeling the spatial information.

Anguera et al \cite{anguera2006beamformit} proposed a Delay-and-Sum beamformer called BeamformIt. The beamformer estimates the weight of each channel and the Time Delay of Arrival (TDOA) between multiple channels and the reference channel by using the GCC-PHAT cross correlation. We apply the speech enhancement with BeamformIt as the baseline.


\section{Methods}
\label{sec:methods}

\subsection{Training the Network to Clean the MESSL Masks}
To improve the quality of the binary masks produced by MESSL, we trained an LSTM neural network to clean a MESSL mask when passing this mask and its associated noisy spectrogram through the network.

The LSTM operates on single-channel recordings.  Each channel in the multi-channel recording is processed independently and in parallel by the LSTM, following work by Erdogan et al. \cite{erdogan2016improved}.  In the single-channel setting, the short-time Fourier transform of the recorded noisy signal, $y(\omega,t)$ is assumed to be
\begin{align}
y(\omega,t) = s(\omega,t) + n(\omega,t)
\end{align}
where $s(\omega,t)$ is the (possibly reverberant) target speech and $n(\omega,t)$ is non-stationary additive noise. In each case the network was configured to output a  $[0,1]$ valued mask $\hat{m}(\omega,t)$ for each frame of the input noisy spectrogram. 

For training targets we used ideal amplitude (IA) masks based on \cite{erdogan2015phase}, defined as:
\begin{align}
m_{ia}(\omega,t) = |s(\omega,t)|/|y(\omega,t)|
\end{align}
The network was trained to minimize the binary cross-entropy loss, using the Nesterov-Adam optimizer \cite{dozat2016incorporating}. Architecturally, the network had 3 bidirectional LTSM layers of 1024 units, followed by a dropout layer for normalization, followed by a dense feedforward layer which outputs the cleaned mask prediction. The LSTM layers merged the bidirectional outputs by averaging them; the dropout rate was set to 0.5; and the dense layer used a sigmoid activation along L2 regularization of the weights.
The spectrogram inputs were converted from a linear to decibel scale, and normalized to mean 0 and variance 1 at each frequency bin. The MESSL binary masks were passed through the logit function. To perform the computation and training of our LSTM neural networks, we used the KERAS python library \cite{chollet2015keras}, built upon the Tensorflow library \cite{tensorflow2015-whitepaper}. We used Keras defaults for any parameter not described above.
We trained the network until the loss on the development set no-longer improved. In practice this occurred around epoch 10.

\begin{figure}[ht!]
    \hspace{-3.0mm}
    \includegraphics[width=0.53\textwidth]{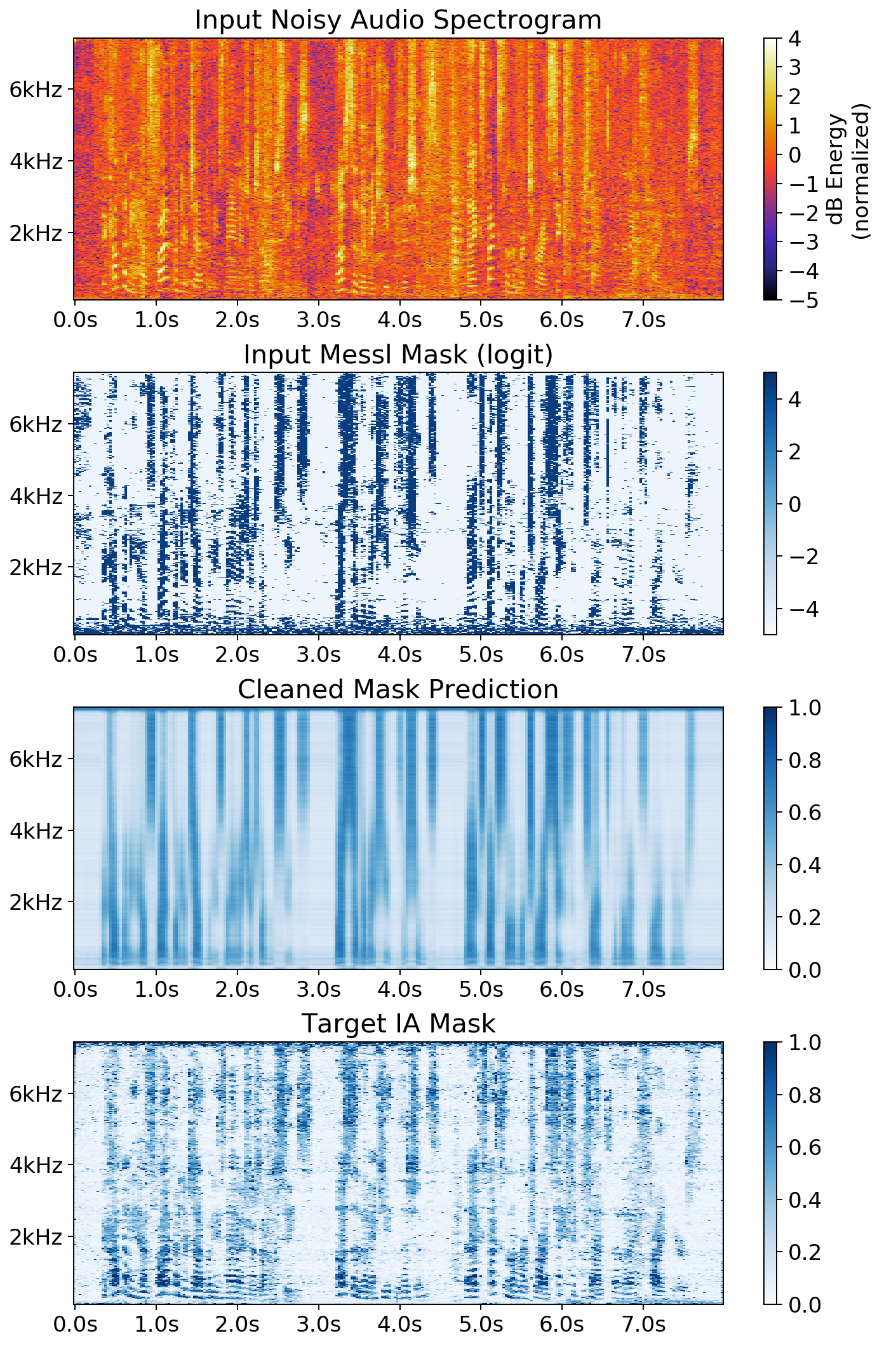}
    \caption{An example of the Mask-Cleaner network using the noisy spectrogram from channel 1 in utterance et05\_bus\_real/F05\_440C0202\_BUS and its MESSL mask to predict an cleaned time-frequency mask.}
    \label{fig:me-ex}
\end{figure}

Figure~\ref{fig:me-ex} gives an example of how our network has learned how to use the noisy spectrogram to clean a mask produced by MESSL.

\subsection{Improved Mask-Driven Beamforming}
A flowchart illustrating the framework of our methods is shown in Figure~\ref{fig:framework}.
We extract six spectrograms from six-channel audio files using short-time Fourier transform (STFT). The window size is 1024 (64ms at 16kHz). We then use one of our models described above to enhance the mask produced by MESSL, using the six different channel spectrograms.

Higuchi et al. \cite{higuchi2016robust} show that using a time-frequency mask to estimate the steering vector helps improve the MVDR beamforming performance. Based on Erdogan et al. \cite{erdogan2016improved}, we propose a slightly modified formula to compute the enhanced spectrogram $\hat{x}(\omega,t)$:
\begin{align}
\hat{x}(\omega,t) =  m_p(\omega,t) \sum_{i=1}^{M}{h_{i}(\omega) y_{i}(\omega,t)} 
\end{align}
where $t$ is the time (frame number), $\omega$ is the frequency, $m_{p}(\omega,t)$ is the time-frequency mask for post processing, $M$ is the number of channels, $y(\omega,t)$ is the noisy spectrogram, and $h_{i}(\omega)$ is the filter for the $i$th channel at frequency $\omega$. 

$h_{i}(\omega)$ is computed by:
\vspace{2mm}
\begin{align}
[h_{1}(\omega), \dots, h_M(\omega)]^T = \frac{1}{\lambda(\omega)}(G(\omega)-I_{M\times M})\times e_{ref}
\end{align}
where $\lambda(\omega) = \Tr(G(\omega))-M$, $e_{ref}$ is the reference microphone unit vector.  We define $G(\omega)$ as:
\vspace{2mm}
\begin{align}
G(\omega) = \Phi^{-1}_{n}(\omega) \times (\Phi_{n}(\omega)+\Phi_{s}(\omega))
\end{align}
where $\Phi_{n}$ and $\Phi_{s}$ are the spatial covariance matrices of the noise and speech respectively, which are computed using separate noise and speech masks $m_{n}(\omega,t)$ and $m_{s}(\omega,t)$.

We compare two different methods for computing the speech, noise, and post-filter masks, first by only using the six LSTM cleaned masks $m_{i}(\omega,t)$ (the outputs of the Mask-Cleaner LSTM network when given a noisy spectrogram and the MESSL mask), then by combining them with the MESSL mask $m_{S}(\omega,t)$.
\vspace{2mm}
\begin{align}
m_{s}(\omega,t) &= \min \left( m_{1}(\omega,t), \dots, m_{M}(\omega,t), m_{S}(\omega,t) \right) \\
m_{n}(\omega,t) &= \max \left( m_{1}(\omega,t), \dots, m_{M}(\omega,t), m_{S}(\omega,t) \right) \\
m_{p}(\omega,t) &= \frac{1}{M+1} \left( m_{S}(\omega,t) + \sum_i m_{i=1}^M(\omega,t) \right)
\end{align}
We combine the masks in this way so that the speech and noise masks are especially conservative in order to accurately estimate the spatial covariances of the speech and noise with minimal leakage from the opposite class. The speech mask only includes points that are definitely speech and the complement of the noise mask only includes points that are definitely noise.  In our results, the first method is referred to as \textbf{LSTM Cleaned Mask}
because $m_S(\omega,t)$ is absent from equations (6)-(8). The second method is referred to as \textbf{MESSL+LSTM} because $m_S(\omega,t)$ is present.




\section{Experiment and Results}
\label{sec:experiments}

\subsection{The CHiME-3 Corpus}
The CHiME-3 corpus features both live and simulated, 6-channel single talker recordings from 12 different talkers (6 male, 6 female), in 4 different noisy environments: caf\'{e}, street junction, public transport and pedestrian area. 
In our work, we used the official data split, with 1600 real and 7138 simulated noisy utterances in the training set for training, along 1640 real and 1640 simulated noisy utterances in the development set for validation.

We tested our models on the proposed 2640 utterances in the test set, which contains audio both from real noisy recordings and simulated noisy recordings.

\subsection{Supervised MVDR Speech Reference}
Because the real subset of the CHiME-3 recordings were spoken in a noisy environment, it is not possible to obtain a true clean reference signal for them for computing the target mask or evaluating enhancement performance.  Instead, an additional microphone was placed close to the talker's mouth to serve as a reference.  While this reference has a higher signal-to-noise ratio than the main microphones, it is not noise free.  In addition, because it is mounted close to the mouth, it contains sounds that are not desired in a clean output and actually could hurt ASR performance, namely pops, lip smacks, and other mouth noises.

In order to obtain a cleaner reference signal, we use the close microphone as a frequency-dependent voice activity detector to control mask-based MVDR beamforming \cite{souden2010} of the main six microphones. The beamformer is adapted in a given frequency when the reference microphone energy is greater than the 85th percentile in that frequency band over an entire utterance.  A final post-filter mask is derived in a similar way, but using the 75th percentile, and then applied with a maximum suppression of 15\,dB. By not including the close microphone signal directly in the estimate, we eliminate the recording artifacts that it contains, while still taking advantage of the information on voice activity that it contains. This is in contrast to other approaches to estimate a reference signal \cite{vincent2016analysis}, which learn a linear transformation to project the close microphone signal onto the microphone array. Such an approach can compensate for different filtering of the speech signal at the close mic and the main mic, but cannot eliminate recording artifacts.

\subsection{Evaluation Metrics}
We evaluate the performance of our enhancement system in terms of both speech quality and intelligibility to a speech recognizer. For quality, we use the Perceptual Evaluation of Speech Quality (PESQ) score \cite{Loizou2007}. PESQ is measured in units of mean opinion score (MOS) between 0 and 5, higher being better. For the simulated data, the reference signals were given by the close mic (CH0) signal from the CHiME-3 booth recordings that were used to make some of the simulated mixtures. For the real data, the reference signals were given by the supervised MVDR for the target speech using the close mic. 
PESQ is fairly accurate at predicting subjective quality scores for speech enhancement, but has the advantage for CHiME-3 of not requiring a reference for the noise sources. 

We also evaluate the enhanced speech by Word Error Rate (WER) using the baseline recipe for CHiME-3 in the Kaldi automatic speech recognition toolkit. We train our Kaldi recognizer on the CHiME-3 corpus. First, we apply BeamformIt to get the single-channel enhanced training data and development data. BeamformIt uses all channels except for channel 2 because channel 2 doesn't face the speaker, thus there is little speech in it. Second, we train a DNN acoustic model using Mel-frequency cepstral coefficients (MFCC) features with feature space maximum likelihood linear regression (fMLLR) speaker adaptation. The DNN model is trained with sMBR criterion according to \cite{vesely2013sequence}. Then we train a 5-gram language model with the CHiME-3 transcripts.

\subsection{Results}
Table~\ref{tab:res1} compares the performance of our systems to the BeamformIt baseline. We also report the PESQ and WER scores of the original CHiME-3 noisy audio on the fifth microphone (Noisy CH5), and audio enhanced by the MESSL system alone (Vanilla MESSL). Our first system (LSTM Cleaned Mask) uses the output of the LSTM alone to perform the beamforming, while our best performing system (MESSL+LSTM) uses the combination of the MESSL mask and the LSTM output.
These scores are computed over the CHiME-3 test dataset.

\begin{table}
\caption{Performance using Perceptual Evaluation of Speech Quality (PESQ, higher scores are better) and automatic speech recognition word error rate (WER, lower scores are better) on CHiME-3 test data. }
\label{tab:res1}
\centering
\begin{tabular}{lll}
\toprule
 & PESQ & WER \\ 
\midrule
Noisy CH5 & 2.05 & 20.0 \\
Vanilla MESSL & 2.16 & 26.9 \\
BeamformIt & 2.19 & 13.6 \\
LSTM Cleaned Mask & 2.54 & 13.5 \\
MESSL+LSTM & \textbf{2.80} & \textbf{10.2}  \\
\bottomrule
\end{tabular}
\end{table}

Table~\ref{tab:res2} details the performance of our best model over all the different data sets provided by CHiME-3, separated by location (bus, cafe, pedestrian area, and street junction) and real vs simulated audio over both the validation and testing CHiME-3 data sets	.

\begin{table}
\caption{Details of the MESSL+LSTM system's performance. }
\label{tab:res2}
\centering
\begin{tabular}{llrrrr}
\toprule
  & & \multicolumn{2}{c}{PESQ} & \multicolumn{2}{c}{WER} \\
  & & dt05 & et05 & dt05 & et05 \\
\midrule
REAL & BUS & 2.52 & 2.55 & 7.42 & 17.14 \\ 
  & CAF & 2.72 & 2.53 & 4.73 & 7.86\\ 
  & PED & 2.82 & 2.41 & 4.44 & 18.61\\ 
  & STR & 2.81 & 2.61 & 6.43 & 9.71 \\ 
SIMU & BUS & 2.99 & 3.01 & 4.38 & 5.83 \\ 
  & CAF & 2.76 & 2.80 & 6.08 & 6.69\\ 
  & PED & 2.93 & 2.81 & 4.99 & 6.99 \\ 
  & STR & 2.78 & 2.81 & 6.40 & 8.87 \\ 
\midrule
Avg &  & 2.07 & 2.80 & 5.61 & 10.20 \\
\bottomrule
\end{tabular}
\end{table}

We also computed the statistical significance of our best model's performance compared to BeamformIt.
\newpage
For PESQ, a paired t-test showed a significant difference ($t = -90.2$, $p < 10^{-5}$).  For WER, a binomial test showed a significant difference ($p < 10^{-5}$). 

\section{Conclusions and Future Work}
\label{sec:conclusion}
In this paper we propose a novel method to adapt parallel single-channel LSTM-based enhancement to multi-channel audio, combining the speech-signal modeling power of the LSTM neural network with the spatial clustering power of MESSL, further enhancing the audio. We show that this method can help MESLL improve the quality of audio, with similar intelligibility.

Our future work will continue to explore different ways of integrating the LSTM speech-signal model with MESSL, in particular integrating the mask cleaning LSTM model in each loop of MESSL's EM algorithm, i.e use the LSTM model to clean the MESSL masks before the estimation of the spatial parameters.
We would also like to explore other training targets, as Erdogan et al. \cite{erdogan2015phase} report improved results when using phase sensitive targets, as well as training directly on the noise-free spectrograms.
%
Additionally the differences in location scores, i.e. the WER performance over the ET05\_BUS\_REAL and ET05\_PED\_REAL sets is significantly worse than the other categories, suggest that more fine tuning of the model is possible. 

\section{Acknowledgements}
This material is based upon work supported by the Alfred P Sloan foundation and the National Science Foundation (NSF) under Grant No. IIS-1409431. Any opinions, findings, and conclusions or recommendations expressed in this material are those of the author(s) and do not necessarily reflect the views of the NSF.

\newpage

\bibliographystyle{IEEEtran}
\bibliography{mybib.bib}

\begin{thebibliography}{10}
\providecommand{\url}[1]{#1}
\csname url@samestyle\endcsname
\providecommand{\newblock}{\relax}
\providecommand{\bibinfo}[2]{#2}
\providecommand{\BIBentrySTDinterwordspacing}{\spaceskip=0pt\relax}
\providecommand{\BIBentryALTinterwordstretchfactor}{4}
\providecommand{\BIBentryALTinterwordspacing}{\spaceskip=\fontdimen2\font plus
\BIBentryALTinterwordstretchfactor\fontdimen3\font minus
  \fontdimen4\font\relax}
\providecommand{\BIBforeignlanguage}[2]{{%
\expandafter\ifx\csname l@#1\endcsname\relax
\typeout{** WARNING: IEEEtran.bst: No hyphenation pattern has been}%
\typeout{** loaded for the language `#1'. Using the pattern for}%
\typeout{** the default language instead.}%
\else
\language=\csname l@#1\endcsname
\fi
#2}}
\providecommand{\BIBdecl}{\relax}
\BIBdecl

\bibitem{achieving-human-parity-conversational-speech-recognition-2}
\BIBentryALTinterwordspacing
W.~Xiong, J.~Droppo, X.~Huang, F.~Seide, M.~Seltzer, A.~Stolcke, D.~Yu, and
  G.~Zweig, ``Achieving human parity in conversational speech recognition,''
  Tech. Rep., February 2017. [Online]. Available:
  \url{https://www.microsoft.com/en-us/research/publication/achieving-human-parity-conversational-speech-recognition-2/}
\BIBentrySTDinterwordspacing

\bibitem{MandelEtAl2017}
M.~I. Mandel, S.~Araki, and T.~Nakatani, ``Multichannel clustering and
  classification approaches,'' in \emph{Audio Source Separation and Speech
  Enhancement}, E.~Vincent, T.~Virtanen, and S.~Gannot, Eds.\hskip 1em plus
  0.5em minus 0.4em\relax Wiley, 2017, ch.~12, to appear.

\bibitem{brandstein2013microphone}
M.~Brandstein and D.~Ward, \emph{Microphone arrays: signal processing
  techniques and applications}.\hskip 1em plus 0.5em minus 0.4em\relax Springer
  Science \& Business Media, 2013.

\bibitem{hochreiter1997long}
S.~Hochreiter and J.~Schmidhuber, ``Long short-term memory,'' \emph{Neural
  computation}, vol.~9, no.~8, pp. 1735--1780, 1997.

\bibitem{6638947}
A.~Graves, A.~r.~Mohamed, and G.~Hinton, ``Speech recognition with deep
  recurrent neural networks,'' in \emph{2013 IEEE International Conference on
  Acoustics, Speech and Signal Processing}, May 2013, pp. 6645--6649.

\bibitem{weninger2015speech}
F.~Weninger, H.~Erdogan, S.~Watanabe, E.~Vincent, J.~Le~Roux, J.~R. Hershey,
  and B.~Schuller, ``Speech enhancement with lstm recurrent neural networks and
  its application to noise-robust asr,'' in \emph{International Conference on
  Latent Variable Analysis and Signal Separation}.\hskip 1em plus 0.5em minus
  0.4em\relax Springer, 2015, pp. 91--99.

\bibitem{xiao16}
X.~Xiao, S.~Watanabe, H.~Erdogan, L.~Lu, J.~Hershey, M.~L. Seltzer, G.~Chen,
  Y.~Zhang, M.~Mandel, and D.~Yu, ``Deep beamforming networks for multi-channel
  speech recognition,'' in \emph{Proceedings of the {IEEE} International
  Conference on Acoustics, Speech, and Signal Processing}.\hskip 1em plus 0.5em
  minus 0.4em\relax IEEE, mar 2016, pp. 5745--5749.

\bibitem{SainathEtAl2015}
T.~N. Sainath, R.~J. Weiss, A.~Senior, K.~W. Wilson, and O.~Vinyals, ``Learning
  the speech front-end with raw waveform cldnns,'' 2015.

\bibitem{5200357}
M.~I. Mandel, R.~J. Weiss, and D.~P.~W. Ellis, ``Model-based
  expectation-maximization source separation and localization,'' \emph{IEEE
  Transactions on Audio, Speech, and Language Processing}, vol.~18, no.~2, pp.
  382--394, Feb 2010.

\bibitem{sawada2011underdetermined}
H.~Sawada, S.~Araki, and S.~Makino, ``Underdetermined convolutive blind source
  separation via frequency bin-wise clustering and permutation alignment,''
  \emph{IEEE Transactions on Audio, Speech, and Language Processing}, vol.~19,
  no.~3, pp. 516--527, 2011.

\bibitem{bagchi15}
\BIBentryALTinterwordspacing
D.~Bagchi, M.~I. Mandel, Z.~Wang, Y.~He, A.~Plummer, and E.~Fosler-Lussier,
  ``Combining spectral feature mapping and multi-channel model-based source
  separation for noise-robust automatic speech recognition,'' in
  \emph{Proceedings of the {IEEE} Workshop on Automatic Speech Recognition and
  Understanding}, 2015. [Online]. Available:
  \url{http://m.mr-pc.org/work/asru15.pdf}
\BIBentrySTDinterwordspacing

\bibitem{rix2001perceptual}
A.~W. Rix, J.~G. Beerends, M.~P. Hollier, and A.~P. Hekstra, ``Perceptual
  evaluation of speech quality (pesq)-a new method for speech quality
  assessment of telephone networks and codecs,'' in \emph{Acoustics, Speech,
  and Signal Processing, 2001. Proceedings.(ICASSP'01). 2001 IEEE International
  Conference on}, vol.~2.\hskip 1em plus 0.5em minus 0.4em\relax IEEE, 2001,
  pp. 749--752.

\bibitem{Povey_ASRU2011}
D.~Povey, A.~Ghoshal, G.~Boulianne, L.~Burget, O.~Glembek, N.~Goel,
  M.~Hannemann, P.~Motlicek, Y.~Qian, P.~Schwarz, J.~Silovsky, G.~Stemmer, and
  K.~Vesely, ``The kaldi speech recognition toolkit,'' in \emph{IEEE 2011
  Workshop on Automatic Speech Recognition and Understanding}.\hskip 1em plus
  0.5em minus 0.4em\relax IEEE Signal Processing Society, Dec. 2011, iEEE
  Catalog No.: CFP11SRW-USB.

\bibitem{mandel16b}
\BIBentryALTinterwordspacing
M.~I. Mandel and J.~P. Barker, ``Multichannel spatial clustering for robust
  far-field automatic speech recognition in mismatched conditions,'' in
  \emph{Proceedings of Interspeech}, 2016, pp. 1991--1995. [Online]. Available:
  \url{http://m.mr-pc.org/work/interspeech16b.pdf}
\BIBentrySTDinterwordspacing

\bibitem{anguera2006beamformit}
X.~Anguera, ``Beamformit, the fast and robust acoustic beamformer,'' 2006.

\bibitem{nugraha2016multichannel}
A.~A. Nugraha, A.~Liutkus, and E.~Vincent, ``Multichannel audio source
  separation with deep neural networks,'' \emph{IEEE/ACM Transactions on Audio,
  Speech, and Language Processing}, vol.~24, no.~9, pp. 1652--1664, 2016.

\bibitem{pfeifenberger2017}
\BIBentryALTinterwordspacing
L.~Pfeifenberger, M.~Zohrer, and F.~Pernkopf, ``Dnn-based speech mask
  estimation for eigenvector beamforming,'' in \emph{Acoustics, Speech and
  Signal Processing (ICASSP), 2017 IEEE International Conference on}.\hskip 1em
  plus 0.5em minus 0.4em\relax IEEE SigPort, 2017. [Online]. Available:
  \url{http://sigport.org/1583}
\BIBentrySTDinterwordspacing

\bibitem{heymann2016neural}
J.~Heymann, L.~Drude, and R.~Haeb-Umbach, ``Neural network based spectral mask
  estimation for acoustic beamforming,'' in \emph{Acoustics, Speech and Signal
  Processing (ICASSP), 2016 IEEE International Conference on}.\hskip 1em plus
  0.5em minus 0.4em\relax IEEE, 2016, pp. 196--200.

\bibitem{heymann2017}
J.~Heymann, L.~Drude, C.~Boeddeker, P.~Hanebrink, and R.~Haeb-Umbach,
  ``Beamnet: End-to-end training of a beamformer-supported multi-channel asr
  system,'' in \emph{Acoustics, Speech and Signal Processing (ICASSP), 2017
  IEEE International Conference on}, 2017.

\bibitem{boeddeker2017}
C.~Boeddeker, P.~Hanebrink, J.~Heymann, D.~Lukas, and R.~Haeb-Umbach,
  ``Optimizing neural-network supported acoustic beamforming by algorithmic
  differentiation,'' in \emph{Acoustics, Speech and Signal Processing (ICASSP),
  2017 IEEE International Conference on}, 2017.

\bibitem{erdogan2016improved}
H.~Erdogan, J.~R. Hershey, S.~Watanabe, M.~Mandel, and J.~Le~Roux, ``Improved
  mvdr beamforming using single-channel mask prediction networks,'' in
  \emph{Proc. INTERSPEECH}, 2016.

\bibitem{erdogan2015phase}
H.~Erdogan, J.~R. Hershey, S.~Watanabe, and J.~Le~Roux, ``Phase-sensitive and
  recognition-boosted speech separation using deep recurrent neural networks,''
  in \emph{Acoustics, Speech and Signal Processing (ICASSP), 2015 IEEE
  International Conference on}.\hskip 1em plus 0.5em minus 0.4em\relax IEEE,
  2015, pp. 708--712.

\bibitem{dozat2016incorporating}
T.~Dozat, ``Incorporating nesterov momentum into adam,'' 2016.

\bibitem{chollet2015keras}
F.~Chollet, ``Keras,'' \url{https://github.com/fchollet/keras}, 2015.

\bibitem{tensorflow2015-whitepaper}
\BIBentryALTinterwordspacing
M.~Abadi and et~al, ``{TensorFlow}: Large-scale machine learning on
  heterogeneous systems,'' 2015, software available from tensorflow.org.
  [Online]. Available: \url{http://tensorflow.org/}
\BIBentrySTDinterwordspacing

\bibitem{higuchi2016robust}
T.~Higuchi, N.~Ito, T.~Yoshioka, and T.~Nakatani, ``Robust mvdr beamforming
  using time-frequency masks for online/offline asr in noise,'' in
  \emph{Acoustics, Speech and Signal Processing (ICASSP), 2016 IEEE
  International Conference on}.\hskip 1em plus 0.5em minus 0.4em\relax IEEE,
  2016, pp. 5210--5214.

\bibitem{souden2010}
M.~Souden, J.~Benesty, and S.~Affes, ``On optimal frequency-domain multichannel
  linear filtering for noise reduction,'' vol.~18, no.~2, pp. 260--276, 2010.

\bibitem{vincent2016analysis}
E.~Vincent, S.~Watanabe, A.~A. Nugraha, J.~Barker, and R.~Marxer, ``An analysis
  of environment, microphone and data simulation mismatches in robust speech
  recognition,'' \emph{Computer Speech \& Language}, 2016.

\bibitem{Loizou2007}
P.~C. Loizou, \emph{Speech Enhancement: Theory and Practice (Signal Processing
  and Communications)}.\hskip 1em plus 0.5em minus 0.4em\relax CRC, 2007.

\bibitem{vesely2013sequence}
K.~Vesel{\`y}, A.~Ghoshal, L.~Burget, and D.~Povey, ``Sequence-discriminative
  training of deep neural networks.'' in \emph{Interspeech}, 2013, pp.
  2345--2349.

\end{thebibliography}

\end{document}